\documentclass[aps,prd,onecolumn,groupedaddress,showpacs,nofootinbib,amssymb]{revtex4}
\usepackage[dvips]{graphicx}
\usepackage{amssymb}
\usepackage{amsmath}
\usepackage{graphicx,,color}
\usepackage{amsfonts}
\usepackage{bm}
\usepackage{cancel}
\usepackage{comment}

\newcommand\be{\begin{equation}}
\newcommand\ee{\end{equation}}

\allowdisplaybreaks[4]

\begin{document}

\tolerance=5000

\title{Revisiting Einstein-Gauss-Bonnet Theories after GW170817}
\author{V.K. Oikonomou$^{1,2}$}
\email{Corresponding author:
voikonomou@gapps.auth.gr;v.k.oikonomou1979@gmail.com}
\affiliation{$^{1)}$Department of Physics, Aristotle University of
Thessaloniki, Thessaloniki 54124, Greece} \affiliation{$^{2)}$L.N.
Gumilyov Eurasian National University - Astana, 010008,
Kazakhstan}

 \tolerance=5000

\begin{abstract}
In this work we study the inflationary framework of
Einstein-Gauss-Bonnet theories which produce a primordial
gravitational wave speed that respects the constraint imposed by
the GW170817 event, namely $\left| c_T^2 - 1 \right| < 6 \times
10^{-15}$ in natural units. For these theories in general, the
scalar Gauss-Bonnet coupling function $\xi (\phi)$ is arbitrary
and unrelated to the scalar potential. We develop the inflationary
formalism of this theory, and present analytic forms for the
slow-roll indices, the observational indices and the $e$-foldings
number, assuming solely a slow-roll era and also that the
constraints imposed by the GW170817 event hold true. We present in
detail an interesting class of models that produces a viable
inflationary era. Finally, we investigate the behavior of the
Einstein-Gauss-Bonnet coupling function during the reheating era,
in which case the evolution of the Hubble rate satisfies a
constant-roll-like relation $\dot{H}=\delta H^2$. As we show, the
behavior of the scalar Gauss-Bonnet coupling during the reheating
is different from that of the inflationary era. This is due to the
fact that the evolution governing the reheating is different
compared to the inflationary era.
\end{abstract}


\maketitle

\section{Introduction}

The inflationary paradigm
\cite{inflation1,inflation2,inflation3,inflation4} for the early
Universe evolution is one exceptional theoretical prediction which
explains most of the shortcomings of the Big Bang paradigm and
also predicts how the large scale structure of our Universe
emerged from the primordial curvature perturbations. To date no
sign of inflation has been detected though, and it is vital for
the validity of the theory to be observationally confirmed. The
next decade is expected to answer the question whether the
inflationary era occurred or not, or at least further constrain
inflationary theories and give signs or hints that inflation
occurred. Indeed, the stage four cosmic microwave background (CMB)
experiments  that are expected to commence in 2027
\cite{CMB-S4:2016ple,SimonsObservatory:2019qwx}, will directly
probe the $B$-modes in the CMB polarization pattern, and the
future gravitational wave experiments
\cite{Hild:2010id,Baker:2019nia,Smith:2019wny,Crowder:2005nr,Smith:2016jqs,Seto:2001qf,Kawamura:2020pcg,Bull:2018lat,LISACosmologyWorkingGroup:2022jok}
will probe inflation indirectly via the stochastic gravitational
wave background, which is believed to have been generated during
the inflationary era. The recent NANOGrav results
\cite{NANOGrav:2023gor} have imposed stringent bounds and
constraints on the inflationary era, and it seems highly unlikely
that a standard inflationary era can explain the NANOGrav 2023
stochastic gravitational wave background \cite{Vagnozzi:2023lwo},
however non-standard inflationary scenarios can potentially
predict an enhanced inflationary spectrum, see for example
\cite{Yi:2023mbm,Balaji:2023ehk}. Thus the future experiments are
expected to shed light on the question whether inflation actually
occurred or not.

Among the candidate theories that can yield an interesting
inflationary era are Einstein-Gauss-Bonnet theories
\cite{Hwang:2005hb,Nojiri:2006je,Cognola:2006sp,Nojiri:2005vv,Nojiri:2005jg,Satoh:2007gn,Bamba:2014zoa,Yi:2018gse,Guo:2009uk,Guo:2010jr,Jiang:2013gza,vandeBruck:2017voa,Pozdeeva:2020apf,Vernov:2021hxo,Pozdeeva:2021iwc,Fomin:2020hfh,DeLaurentis:2015fea,Chervon:2019sey,Nozari:2017rta,Odintsov:2018zhw,Kawai:1998ab,Yi:2018dhl,vandeBruck:2016xvt,Maeda:2011zn,Ai:2020peo,Easther:1996yd,Codello:2015mba},
which can predict a blue-tilted tensor spectrum. In general,
Einstein-Gauss-Bonnet theories are strongly motivated since these
are basically string-corrected scalar field theories. Let us
recall that the most general scalar field Lagrangian in four
dimensions, that contains two derivatives  at most is,
\begin{equation}\label{generalscalarfieldaction}
\mathcal{S}_{\varphi}=\int
\mathrm{d}^4x\sqrt{-g}\left(\frac{1}{2}Z(\varphi)g^{\mu
\nu}\partial_{\mu}\varphi
\partial_{\nu}\varphi+\mathcal{V}(\varphi)+h(\varphi)\mathcal{R}
\right)\, ,
\end{equation}
and by evaluating the scalar field at its vacuum configuration, it
must be either minimally or conformally coupled. By considering
the former case, thus $Z(\varphi)=-1$ and also $h(\varphi)=1$ in
the action (\ref{generalscalarfieldaction}), the most general
quantum corrections containing up to fourth order derivatives, of
the local effective action, consistent with diffeomorphism
invariance, is \cite{Codello:2015mba},
\begin{align}\label{quantumaction}
&\mathcal{S}_{eff}=\int
\mathrm{d}^4x\sqrt{-g}\Big{(}\Lambda_1+\Lambda_2
\mathcal{R}+\Lambda_3\mathcal{R}^2+\Lambda_4 \mathcal{R}_{\mu
\nu}\mathcal{R}^{\mu \nu}+\Lambda_5 \mathcal{R}_{\mu \nu \alpha
\beta}\mathcal{R}^{\mu \nu \alpha \beta}+\Lambda_6 \square
\mathcal{R}\\ \notag &
+\Lambda_7\mathcal{R}\square\mathcal{R}+\Lambda_8 \mathcal{R}_{\mu
\nu}\square \mathcal{R}^{\mu
\nu}+\Lambda_9\mathcal{R}^3+\mathcal{O}(\partial^8)+...\Big{)}\, ,
\end{align}
with the parameters $\Lambda_i$, $i=1,2,...,6$ being appropriate
dimensionful constants. In 2017 however, the GW170817 event
observed by the LIGO-Virgo collaboration
\cite{TheLIGOScientific:2017qsa,Monitor:2017mdv,GBM:2017lvd,LIGOScientific:2019vic}
altered our perspective on several candidate theories of
inflation, because the observation indicated that the
gravitational waves have a propagation speed that it is almost
equal to that of light. This result has put severe constraints on
theories that predict a propagation speed of tensor perturbations
different from that of light, see Refs.
\cite{Ezquiaga:2017ekz,Baker:2017hug,Creminelli:2017sry,Sakstein:2017xjx,Boran:2017rdn}.
However, the GW170817 event is a late-time event, thus one might
claim that the speed of primordial tensor perturbations could be
different from that of light primordially. This argument however
is not correct because inflation is basically a classical theory
corresponding to a four dimensional post-Planck Universe. Thus
there is no fundamental reason that could make the graviton to
change its on-shell mass during and after the inflationary era.
Hence, the rule in massive gravity theories is that the speed of
tensor perturbations should be equal to that of light. In a series
of articles
\cite{Oikonomou:2021kql,Oikonomou:2022xoq,Odintsov:2020sqy}, the
requirement that the gravitational wave speed is equal to that of
light was modelled by using the constraint $\ddot{\xi}=H\dot{\xi}$
in Einstein-Gauss-Bonnet theories. This constraint made the scalar
potential and the scalar Gauss-Bonnet coupling to be
interdependent. In this work we aim to study the inflationary
phenomenology of Einstein-Gauss-Bonnet theories by using the
actual phenomenological constraints on the gravitational wave
speed imposed by the GW170817 event, which is $\left| c_T^2 - 1
\right| < 6 \times 10^{-15}$ in natural units. Thus, we shall
model inflation in Einstein-Gauss-Bonnet theories without taking
the constraint $\ddot{\xi}=H\dot{\xi}$ into account. In this way,
the scalar Gauss-Bonnet coupling function $\xi (\phi)$ is free to
choose and is not directly related to the potential by some
physically motivated constraint. We develop the formalism of
Einstein-Gauss-Bonnet inflationary theories and we extract
analytic relations for the slow-roll indices and the observational
indices of inflation. Among a large variety of models, we present
a promising class of potentials and scalar Gauss-Bonnet couplings
that yield a viable inflationary phenomenology compatible with the
GW170817 constraint. Finally, we consider the behavior of the
Einstein-Gauss-Bonnet theory during the reheating era, in which
case the Hubble rate and its time derivative satisfy a
constant-roll-like relation $\dot{H}=\delta H^2$. In this case,
the scalar Gauss-Bonnet coupling function is required to obey a
differential equation that relates it directly to the scalar
potential. We examine several potentials which yielded a viable
inflationary phenomenology and we find the functional form of the
scalar Gauss-Bonnet coupling for these potentials, and these are
different from the forms of the scalar Gauss-Bonnet couplings
during the inflationary era.

This article is organized as follows: In section II we develop the
Einstein-Gauss-Bonnet inflation formalism, we extract analytic
forms for the slow-roll indices and the observational indices of
inflation. We also impose several constraints in order to comply
with the GW170817 event. In section III we apply the formalism by
using a class of models that yields a viable phenomenology for
inflationary theories. We determine the range of values of the
free parameters of the theory that produce a viable inflationary
phenomenology and at the same time a gravitational wave speed
which respects the constraint $\left| c_T^2 - 1 \right| < 6 \times
10^{-15}$ in natural units. Also, in section III we study the
behavior of Einstein-Gauss-Bonnet theories during the reheating
era and we extract the functional form of the scalar Gauss-Bonnet
coupling during this era, given the scalar potential. Finally, the
conclusions appear in the end of this article.

\section{Essential Features of Einstein-Gauss-Bonnet in View of GW170817 Event}

We start our analysis by presenting the essential features of
Einstein-Gauss-Bonnet theories in view of the GW170817 event. The
gravitational action of Einstein-Gauss-Bonnet theories is,
\begin{equation}
\label{action} \centering
S=\int{d^4x\sqrt{-g}\left(\frac{R}{2\kappa^2}-\frac{1}{2}\partial_{\mu}\phi\partial^{\mu}\phi-V(\phi)-\frac{1}{2}\xi(\phi)\mathcal{G}\right)}\,
,
\end{equation}
where $R$ denotes as usual the Ricci scalar,
$\kappa=\frac{1}{M_p}$ where $M_p$ is the reduced Planck mass, and
also $\mathcal{G}$ stands for the Gauss-Bonnet invariant in four
dimensions, which in terms of the Ricci scalar, the Ricci tensor
$R_{\alpha\beta}$ and the Riemann tensor
$R_{\alpha\beta\gamma\delta}$ reads as follows
$\mathcal{G}=R^2-4R_{\alpha\beta}R^{\alpha\beta}+R_{\alpha\beta\gamma\delta}R^{\alpha\beta\gamma\delta}$.
We assume that the geometric background is a flat
Friedmann-Lemaitre-Robertson-Walker (FLRW) spacetime of the form,
\begin{equation}
\label{metric} \centering
ds^2=-dt^2+a(t)^2\sum_{i=1}^{3}{(dx^{i})^2}\, .
\end{equation}
We assume that the scalar field is solely time-dependent, so the
variation of the gravitational action (\ref{action}) with respect
to the scalar field and with respect to the metric yields the
following field equations,
\begin{equation}
\label{motion1} \centering
\frac{3H^2}{\kappa^2}=\frac{1}{2}\dot\phi^2+V+12 \dot\xi H^3\, ,
\end{equation}
\begin{equation}
\label{motion2} \centering \frac{2\dot
H}{\kappa^2}=-\dot\phi^2+4\ddot\xi H^2+8\dot\xi H\dot H-4\dot\xi
H^3\, ,
\end{equation}
\begin{equation}
\label{motion3} \centering \ddot\phi+3H\dot\phi+V'+12 \xi'H^2(\dot
H+H^2)=0\, .
\end{equation}
In order for the inflationary era to occur, a requirement is that
$\dot{H}\ll H^2$, and in addition to that we shall assume that the
scalar field slow-rolls its potential during the whole
inflationary era, so the following requirements are assumed to
hold true,
\begin{equation}\label{slowrollhubble}
\dot{H}\ll H^2,\,\,\ \frac{\dot\phi^2}{2} \ll V,\,\,\,\ddot\phi\ll
3 H\dot\phi\, .
\end{equation}
Einstein-Gauss-Bonnet gravities are known to be plagued with the
issue related to the propagation speed of primordial gravitational
waves, or equivalently with the propagation speed of the tensor
perturbations, which is \cite{Hwang:2005hb},
\begin{equation}
\label{GW} \centering c_T^2=1-\frac{Q_f}{2Q_t}\, ,
\end{equation}
where the functions $Q_f$, $F$ and $Q_b$ are equal to $Q_f=8
(\ddot\xi-H\dot\xi)$, $Q_t=F+\frac{Q_b}{2}$,
$F=\frac{1}{\kappa^2}$ and also $Q_b=-8 \dot\xi H$. Now there are
two ways to take into account the constraints imposed by the
GW170817 event, either by requiring that $Q_f=0$, hence one gets
$c_T^2=1$ for all the subsequent eras after inflation, or by
imposing direct constraints on the values of the terms entering in
the gravitational wave speed (\ref{GW}). The first approach was
considered in a series of papers
\cite{Oikonomou:2021kql,Oikonomou:2022xoq,Odintsov:2020sqy} so we
shall not consider this approach in this paper. Instead, we shall
assume that the terms entering in the expression of the
gravitational wave speed, are sufficiently small, so that the
gravitational wave speed satisfies the constraint imposed by the
GW170817 event, which is,
\begin{align}
\label{GWp9} \left| c_T^2 - 1 \right| < 6 \times 10^{-15}\, .
\end{align}
In order to achieve this, we assume that,
\begin{equation}\label{actualgw170817constraints}
\kappa^2\dot{\xi}H\ll 1,\,\,\,\kappa^2\ddot{\xi}\ll 1\, ,
\end{equation}
and if these constraints are satisfied during the inflationary
era, then the gravitational wave speed can be sufficiently small
in order to satisfy the constraint (\ref{GWp9}). Let us highlight
an important issue here having to do with the inequalities
(\ref{actualgw170817constraints}), these are not chosen for the
sake of analytic simplicity, but these basically quantify the
requirement that $c_T^2\sim 1$. Indeed, by looking at Eq. (9) we
can see that the gravitational wave speed deviates from unity once
the parameter $Q_f=8 (\ddot\xi-H\dot\xi)$ deviates from zero. This
parameter can be small enough if $\ddot{\xi}\simeq H\dot{\xi}$, a
case thoroughly analyzed in Refs. [56-58], or if each of the terms
$\kappa^2 \ddot{\xi}$ and $\kappa^2 \dot{\xi}H$ are independently
very small. This is exactly what the inequalities presented in Eq.
(11) signify. Note that the approach we adopted does not constrain
the functional form of the Gauss-Bonnet scalar coupling function
$\xi(\phi)$, which is free to choose, in contrast to the approach
adopted in Refs.
\cite{Oikonomou:2021kql,Oikonomou:2022xoq,Odintsov:2020sqy}, which
constrain the functional form of the Gauss-Bonnet scalar coupling
function $\xi(\phi)$ to be related to the scalar potential. In our
approach both the scalar potential and the Gauss-Bonnet scalar
coupling function are free to choose. Now let us consider the
Friedmann equation, and notice that since we assumed that the
constraints of Eq. (\ref{actualgw170817constraints}) hold true,
the term $H^2$ dominates over the terms $\kappa^2 \dot{\xi}H^3$.
We also make another crucial assumption, regarding the
Raychaudhuri equation, that is, we assume that
$\kappa^2\dot{\xi}H^3\ll \kappa^2\dot{\phi}^2$ and
$\kappa^2\ddot{\xi}H^2\ll \kappa^2\dot{\phi}^2$, that is,
\begin{equation}\label{additionalconstraints}
\kappa^2\dot{\xi}H^3\ll
\kappa^2\dot{\phi}^2,\,\,\,\kappa^2\ddot{\xi}H^2\ll
\kappa^2\dot{\phi}^2\, .
\end{equation}
The constraints (\ref{additionalconstraints}) are additional
constraints, which must be satisfied from any viable model. To sum
up, the slow-roll and inflationary constraints
(\ref{slowrollhubble}) are natural constraints imposed by
inflationary dynamics, the constraints
(\ref{actualgw170817constraints}) are additional constraints
imposed by the GW170817 event, and finally the constraints
(\ref{additionalconstraints}) are rather logical assumptions,
which however must be checked if they hold true for any viable
model we shall develop. Now in view of the above, the field
equations take the simpler form, regarding the Friedmann equation,
it becomes,
\begin{equation}
\label{motion5} \centering H^2\simeq\frac{\kappa^2V}{3}\, ,
\end{equation}
while the Raychaudhuri equation becomes,
\begin{equation}
\label{motion6} \centering \dot H\simeq-\frac{1}{2}\kappa^2
\dot\phi^2\, ,
\end{equation}
and the modified Klein-Gordon equation yields,
\begin{equation}
\label{motion8} \centering \dot\phi\simeq
-\frac{12\xi'(\phi)H^4+V'}{3H}\, .
\end{equation}
Now the inflationary phenomenology can be studied by using the
slow-roll indices and the corresponding observational indices of
inflation. Regarding the slow-roll indices, these have the
following general form \cite{Hwang:2005hb},
\begin{align}
\centering \label{indices} \epsilon_1&=-\frac{\dot
H}{H^2}&\epsilon_2&=\frac{\ddot\phi}{H\dot\phi}&\epsilon_3&=0&\epsilon_4&=\frac{\dot
E}{2HE}&\epsilon_5&=\frac{Q_a}{2HQ_t}&\epsilon_6&=\frac{\dot
Q_t}{2HQ_t}\, ,
\end{align}
where $Q_a=-4\dot{\xi} H^2 $, $Q_b=-8\dot{\xi} H$,
$E=\frac{1}{(\kappa\dot\phi)^2}\left(
\dot\phi^2+\frac{3Q_a^2}{2Q_t}+Q_c\right)$, $Q_c=0$, $Q_d=0$,
$Q_e=-16 \dot{\xi} \dot{H}$, $Q_f=8\left(\ddot{\xi}-\dot{\xi}H
\right)$ and also $Q_t=\frac{1}{\kappa^2}+\frac{Q_b}{2}$.
Regarding the observables of inflation, the scalar spectral index
of the primordial curvature perturbations in terms of the
slow-roll indices, is equal to,
\begin{equation}\label{spectralindex}
n_{\mathcal{S}}=1+\frac{2 (-2
\epsilon_1-\epsilon_2-\epsilon_4)}{1-\epsilon_1}\, ,
\end{equation}
and the tensor-to-scalar ratio is equal to,
\begin{equation}\label{tensortoscalar}
r=\left |\frac{16 \left(c_A^3 \left(\epsilon_1-\frac{1}{4} \kappa
^2 \left(\frac{2
Q_c+Q_d}{H^2}-\frac{Q_e}{H}+Q_f\right)\right)\right)}{c_T^3
\left(\frac{\kappa ^2 Q_b}{2}+1\right)}\right |\, ,
\end{equation}
with $c_A$ being the sound speed of the scalar perturbations which
is,
\begin{equation}\label{soundspeed}
c_A=\sqrt{\frac{\frac{Q_a Q_e}{\frac{2}{\kappa ^2}+Q_b}+Q_f
\left(\frac{Q_a}{\frac{2}{\kappa
^2}+Q_b}\right)^2+Q_d}{\dot{\phi}^2+\frac{3 Q_a^2}{\frac{2}{\kappa
^2}+Q_b}+Q_c}+1}\, ,
\end{equation}
and in addition $c_T$ is the gravitational wave speed appearing in
Eq. (\ref{GW}). Furthermore, the spectral index of the tensor
perturbations is,
\begin{equation}\label{tensorspectralindex}
n_{\mathcal{T}}=-2\frac{\left( \epsilon_1+\epsilon_6
\right)}{1-\epsilon_1}\, .
\end{equation}
An essential and important quantity for the study of the
phenomenological viability of an Einstein-Gauss-Bonnet model is
the $e$-foldings number expressed in terms of known or given
quantities. This is defined as follows,
\begin{equation}
\label{efolds} \centering
N=\int_{t_i}^{t_f}{Hdt}=\int_{\phi_i}^{\phi_f}\frac{H}{\dot{\phi}}d\phi\,
,
\end{equation}
where $\phi_i$ and $\phi_f$ denote the scalar field values at
first horizon crossing, in the beginning of inflation, and the end
of the inflationary era respectively. We can find the value of the
scalar field at the end of inflation easily by equating the first
slow-roll index to unity  $|\epsilon_1|\sim \mathcal{O}(1)$, since
the condition $\dot{H}=-H^2$ indicates the end of inflation, it is
the condition of non-acceleration. Alternatively, if the first
slow-roll index is constant, another condition would be
$\epsilon_2\sim \mathcal{O}(1)$ because this would indicate the
end of the slow-roll era, which is equivalent to the end of
inflation condition. In view of Eqs. (\ref{efolds}) and
(\ref{motion8}), the $e$-foldings number (\ref{efolds}) is written
as follows,
\begin{equation}
\label{efolds1} \centering
N=\int_{\phi_i}^{\phi_f}\frac{2H^2}{12\xi'H^4+V'}d\phi\, ,
\end{equation}
which in view of the Friedmann equation (\ref{motion5}) is finally
written as follows,
\begin{equation}
\label{efoldsuncostrained} \centering
N=\int_{\phi_f}^{\phi_i}\frac{\kappa ^2 V(\phi )}{V'(\phi
)+\frac{4}{3} \kappa ^4 V(\phi )^2 \xi '(\phi )}d\phi\, .
\end{equation}
Another important quantity, necessary for the phenomenological
viability of a model, is the amplitude of scalar perturbations
$\mathcal{P}_{\zeta}(k_*)$ and the constraints on this quantity
imposed by the latest Planck data \cite{Planck:2018jri}. The
amplitude of the scalar perturbations is defined as follows,
\begin{equation}\label{definitionofscalaramplitude}
\mathcal{P}_{\zeta}(k_*)=\frac{k_*^3}{2\pi^2}P_{\zeta}(k_*)\, ,
\end{equation}
evaluated at the first horizon crossing when inflation commenced,
where $k_*$ is the CMB pivot scale. The constraint on the
amplitude of the scalar perturbations coming from the Planck data
is $\mathcal{P}_{\zeta}(k_*)=2.196^{+0.051}_{-0.06}\times 10^{-9}$
\cite{Planck:2018jri} at the CMB pivot scale. We can express the
amplitude of the scalar perturbations $\mathcal{P}_{\zeta}(k)$ in
terms of the two point function of the curvature perturbation
$\zeta (k)$ as follows,
\begin{equation}\label{twopointfunctionforzeta}
\langle\zeta(k)\zeta (k')\rangle=(2\pi)^3 \delta^3 (k-k')
P_{\zeta}(k)\, .
\end{equation}
For Einstein-Gauss-Bonnet theories, the amplitude of the scalar
perturbations $\mathcal{P}_{\zeta}(k)$ expressed in terms of the
slow-roll parameters is given below \cite{Hwang:2005hb},
\begin{equation}\label{powerspectrumscalaramplitude}
\mathcal{P}_{\zeta}(k)=\left(\frac{k \left((-2
\epsilon_1-\epsilon_2-\epsilon_4) \left(0.57\, +\log \left(\left|k
\eta \right| \right)-2+\log (2)\right)-\epsilon_1+1\right)}{(2 \pi
) \left(z c_A^{\frac{4-n_{\mathcal{S}}}{2}}\right)}\right)^2\, ,
\end{equation}
where $z=\frac{a \dot{\phi} \sqrt{\frac{E(\phi )}{\frac{1}{\kappa
^2}}}}{H (\epsilon_5+1)}$ and all the quantities must be evaluated
at the first horizon crossing, and furthermore
$\eta=-\frac{1}{aH}\frac{1}{-\epsilon_1+1}$ at first horizon
crossing \cite{Hwang:2005hb}.

Now, we have all the necessary formulas in order to perform viable
model building for GW170817-compatible Einstein-Gauss-Bonnet
theories, so let us recapitulate here all the necessary formulas
and techniques, and also we discuss the strategy towards choosing
a convenient form for the Gauss-Bonnet scalar coupling function.
One starts by solving the equation
$|\epsilon_1(\phi_f)|=\mathcal{O}(1)$, and the value of the scalar
field is determined. Then one solves the equation (\ref{efolds1})
with respect to $\phi_i$, after the integration is performed, and
accordingly the slow-roll indices and the observational indices of
inflation are evaluated for the value $\phi_i$ of the scalar field
at first horizon crossing.

Regarding the strategy we shall adopt for choosing the appropriate
functional form for the Gauss-Bonnet coupling function $\xi
(\phi)$, the choice will be made based on which one yields
analytically simple results. In the present formalism $\xi (\phi)$
is basically free to choose and it is not constrained in any way,
in contrast to the scenario studied in Refs.
\cite{Oikonomou:2021kql,Oikonomou:2022xoq,Odintsov:2020sqy}. Thus
in the case at hand, the most elegant solution and simple to
extract will be determined by the functional form of the first
slow-roll index $\epsilon_1$ and by the integral in Eq.
(\ref{efoldsuncostrained}). The exact form of the slow-roll index
$\epsilon_1$ in terms of the scalar potential is very simple, so
by using the Friedmann equation (\ref{motion5}), the Raychaudhuri
equation (\ref{motion6}) and the modified Klein-Gordon equation
(\ref{motion8}), we get,
\begin{equation}\label{epsilon1analytic}
\epsilon_1=\frac{4}{3} \kappa ^2 \xi '(\phi ) V'(\phi
)+\frac{V'(\phi )^2}{2 \kappa ^2 V(\phi )^2}+\frac{8}{9} \kappa ^6
V(\phi )^2 \xi '(\phi )^2\, ,
\end{equation}
and the important issue is to find a convenient form for
$\xi'(\phi)$ which can simplify both the functional form of
$\epsilon_1$ appearing in Eq. (\ref{epsilon1analytic}) and also to
simplify the quantity appearing in the integral
(\ref{efoldsuncostrained}). In view of this line of research, in
the next section we shall present several classes of models we did
examine and we analyze the resulting phenomenology in detail, also
examining whether the constraints we assumed hold true.

\section{Phenomenological Viability of Several Classes of Einstein-Gauss-Bonnet Models}

\begin{figure}
\centering
\includegraphics[width=25pc]{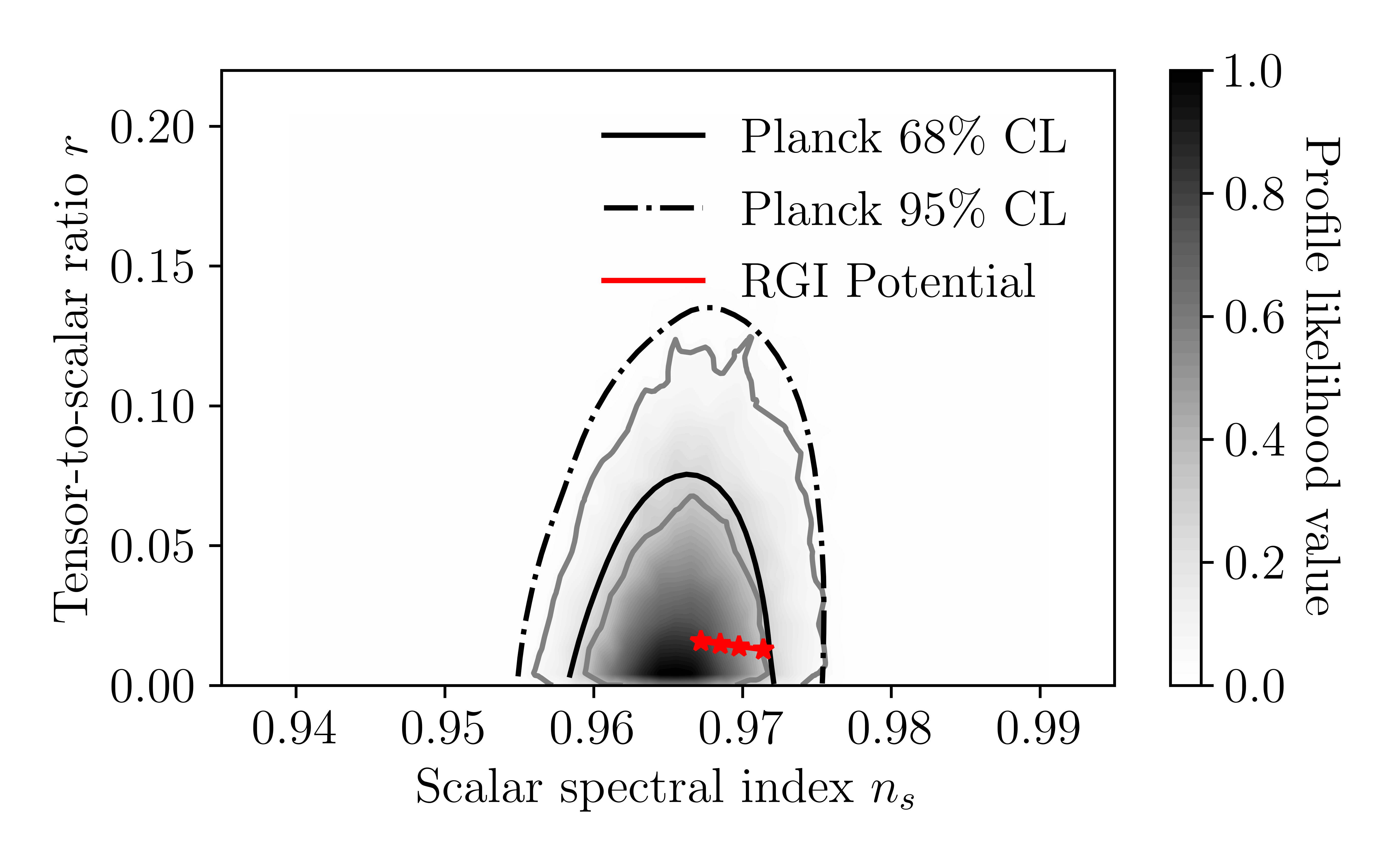}
\caption{Marginalized curves of the Planck 2018 data and the RGI
potential inflation predictions (red curve).}\label{plot1}
\end{figure}

\begin{figure}
\centering
\includegraphics[width=25pc]{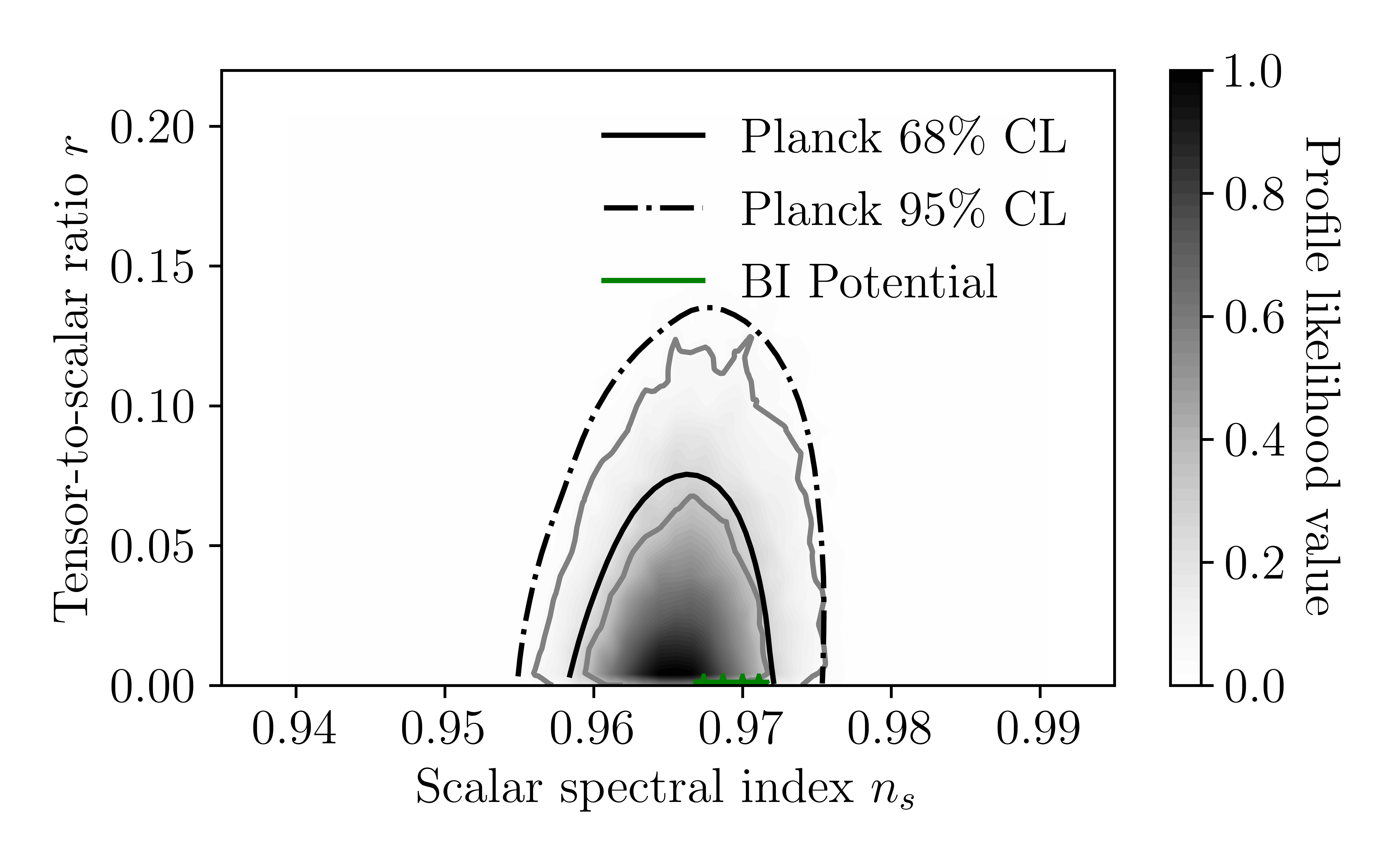}
\caption{Marginalized curves of the Planck 2018 data and the BI
potential inflation predictions (green curve).}\label{plot2}
\end{figure}

\begin{figure}
\centering
\includegraphics[width=25pc]{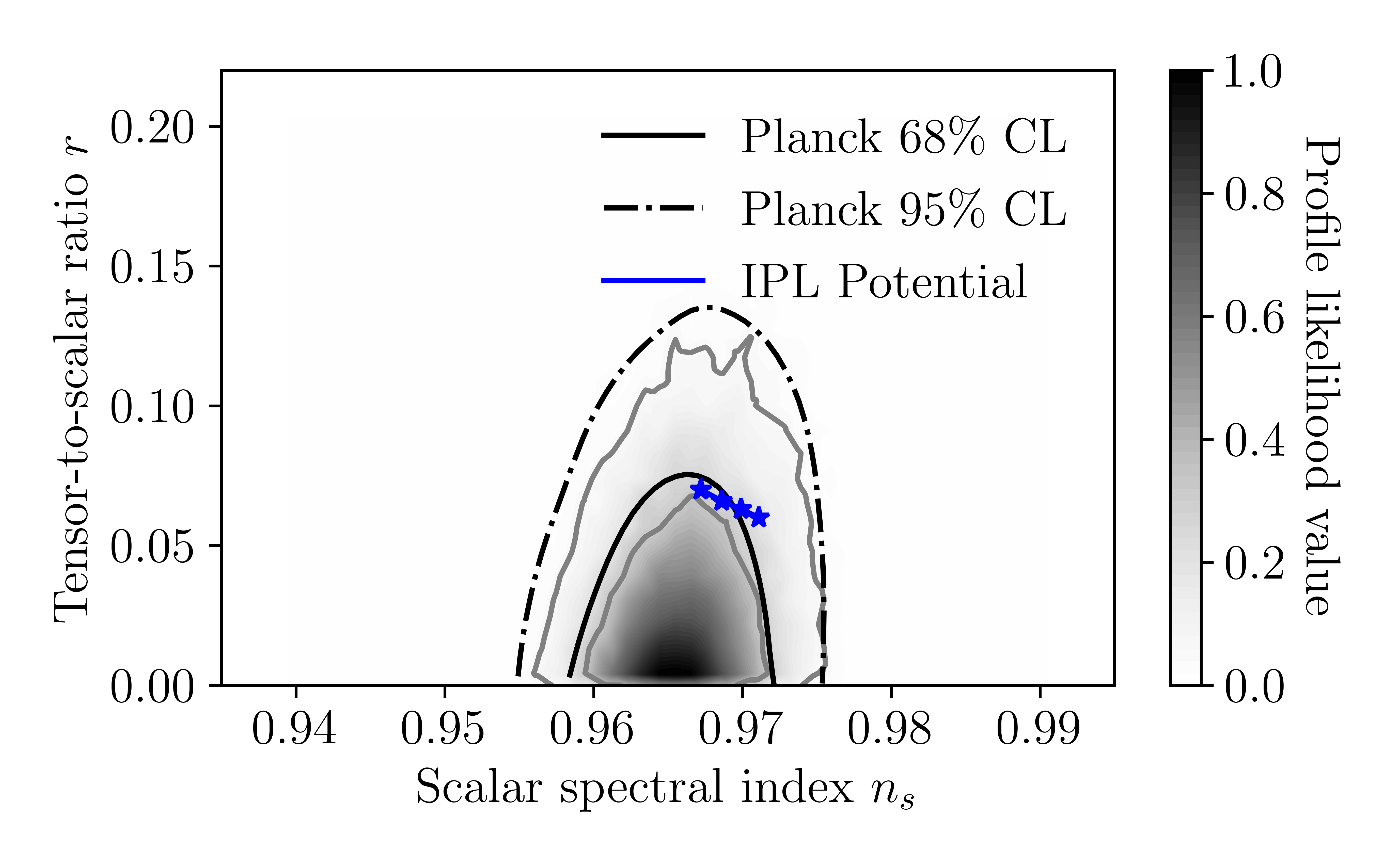}
\caption{Marginalized curves of the Planck 2018 data and the IPL
potential inflation predictions (blue curve).}\label{plot3}
\end{figure}

In this section we shall consider several viable models of
Einstein-Gauss-Bonnet gravity which yield both a viable
inflationary era and also produce a propagation speed of tensor
perturbations that is almost equal to the speed of light and
satisfy the constraint (\ref{GWp9}). For convenience, we shall use
the Planck units system for which,
$$\kappa^2=\frac{1}{8\pi G}=\frac{1}{M_P^2}=1\, .$$ Among the
numerous choices for the Gauss-Bonnet coupling function which can
be made, the most viable scenarios and the most easy to tackle
analytically are obtained by using the following choice,
\begin{equation}\label{couplingfunctionchoices3}
\xi'(\phi)=\frac{\lambda  V'(\phi )}{V(\phi )^2}\, .
\end{equation}
In this case, the propagation speed of the tensor perturbations
acquires a simple form, which is,
\begin{equation}\label{gwspeedclass2}
c_T^2=\frac{-8 \lambda  (4 \lambda +3)^2 V(\phi ) V'(\phi )^2
V''(\phi )+10 \lambda  (4 \lambda +3)^2 V'(\phi )^4+27 V(\phi
)^4}{3 V(\phi )^2 \left(4 \lambda  (4 \lambda +3) V'(\phi )^2+9
V(\phi )^2\right)}\, ,
\end{equation}
and the first slow-roll index takes the form,
\begin{equation}\label{epsilon1formclass}
\epsilon_1=\frac{(4 \lambda +3)^2 V'(\phi )^2}{18 V(\phi )^2}\, ,
\end{equation}
while the $e$-foldings number integral takes the form,
\begin{equation}\label{finalinitialefoldings}
N=\int_{\phi_f}^{\phi_i} \frac{V(\phi )}{\frac{4}{3} \lambda
V'(\phi )+V'(\phi )} \mathrm{d}\phi\, .
\end{equation}
Also the tensor-to-scalar ratio takes the form,
\begin{equation}\label{tensorspectralindexclass}
n_{\mathcal{T}}=-\frac{2 (4 \lambda +3)^2 V'(\phi )^2 \left(4
\lambda  (4 \lambda +9) V'(\phi )^2-24 \lambda  V(\phi ) V''(\phi
)+9 V(\phi )^2\right)}{\left(4 \lambda  (4 \lambda +3) V'(\phi
)^2+9 V(\phi )^2\right) \left(18 V(\phi )^2-(4 \lambda +3)^2
V'(\phi )^2\right)}\, ,
\end{equation}
while the rest of the observational indices and the slow-roll
parameters have quite complicated form to present these here. Now,
there are many potentials that can generate a viable evolution,
which we list here,
\begin{equation}\label{viablepotentials1}
V(\phi)=M \left(1-\frac{\delta }{\kappa  \phi
}\right),\,\,\,V(\phi)=M \left(1-\frac{\delta }{\kappa  \phi
^2}\right) ,\,\,\, V(\phi)=M (1-d (\kappa  \phi ))^2\, ,
\end{equation}
\begin{equation}\label{viablepotentials2}
V(\phi)=M \left(1-\frac{d}{\kappa  \phi
}\right)^2,\,\,\,V(\phi)=\frac{M}{1-\frac{d}{\kappa  \phi }}
,\,\,\, V(\phi)=M \sqrt{1-\frac{d}{\kappa  \phi }}\, ,
\end{equation}
\begin{equation}\label{viablepotentials3}
V(\phi)=\frac{M (\kappa  \phi )^2}{d+(\kappa  \phi
)^2},\,\,\,V(\phi)=M \sqrt[100]{1-\frac{d}{\kappa  \phi }} ,\,\,\,
V(\phi)=M \left(1-\frac{d}{\kappa  \phi }\right)^{3/2}
\end{equation}
\begin{equation}\label{viablepotentials4}
V(\phi)=M \sqrt[5]{1-\frac{d}{\kappa  \phi }}\, .
\end{equation}
From the above, the last one, namely (\ref{viablepotentials4}) is
viable only for a short inflationary era with $N\sim 35$
$e$-foldings, so let us consider three examples from the above
which are viable for $N\sim 50$ $e$-foldings and also have some
physical significance in single field scalar field theories.
Consider first the potential,
\begin{equation}\label{potential}
V(\phi)=\frac{M (\kappa  \phi )^2}{d+(\kappa  \phi )^2}\, ,
\end{equation}
which is known in single scalar field theory as radion gauge
inflation scalar potential (RGI), developed in Ref.
\cite{Fairbairn:2003yx}. In this case, the analytic form of the
derivative of the Gauss-Bonnet scalar coupling function in Planck
units is,
\begin{equation}\label{xiphi}
\xi'(\phi)=\frac{2 d \lambda }{M \phi ^3}\, ,
\end{equation}
hence the first slow-roll index takes the form,
\begin{equation}\label{slowrollindexena}
\epsilon_1=\frac{2 d^2 (4 \lambda +3)^2}{9 \phi ^2 \left(d+\phi
^2\right)^2}\, ,
\end{equation}
and therefore one may solve the equation $|\epsilon_1(\phi_f)|=1$
and obtain $\phi_f$, but we do not quote it here because it is too
lengthy. In this case, the integral involved in the computation of
the $e$-foldings number is quite easy to obtain, and it is equal
to,
\begin{equation}\label{integraln1}
N=\frac{\frac{3 d \phi ^2}{2}+\frac{3 \phi ^4}{4}}{8 d \lambda +6
d}\Big{|}_{\phi_f}^{\phi_i}\, ,
\end{equation}
and from it, the value of the scalar field at the beginning of
inflation, namely $\phi_i$ can be obtained. A viable phenomenology
is obtained for $N=50$ for various choices of the free parameters
$(d,\lambda,M)$, for example for
$(d,\lambda,M)=(3,10^{-13},2.583\times 10^{-10})$, in which case
we get, $n_{\mathcal{S}}=0.96854$, $r= 0.0148501$ and
$n_{\mathcal{T}}=-0.000928563$, hence the model is viable
regarding the observational indices of inflation and well
compatible with the Planck data. In Fig. \ref{plot1} we plot the
model's predictions versus the Planck 2018 likelihood curves, for
$N=[48,55]$ and it is apparent that the model is viable.
Regarding the amplitude of the scalar perturbations for the
current model, for the choice
$(d,\lambda,M)=(3,10^{-13},2.583\times 10^{-10})$ we get exactly
$\mathcal{P}_{\zeta}(k_*)=2.196\times 10^{-9}$, and the amplitude
of the scalar perturbations crucially depends on the value of $M$
as expected. For $(d,\lambda,M)=(3,10^{-13},2.583\times 10^{-10})$
we obtain super-Planckian values for the scalar field at the
beginning and the end of inflation, but this is a model dependent
feature as we show later using other models. For
$(d,\lambda,M)=(3,10^{-13},2.583\times 10^{-10})$ we get,
$(\phi_i,\phi_f)=(5.64629,1.03964)$. Regarding the gravitational
wave speed, we have $c_T^2-1=-1.19917\times 10^{-16}$, so the
model is consistent with the constraint (\ref{GWp9}). Regarding
the constraints (\ref{additionalconstraints}), these are well
satisfied since for $(d,\lambda,M)=(3,10^{-13},2.583\times
10^{-10})$ we have $\dot{\phi}^2\sim\mathcal{O}(10^{-14})$,
$\dot{\xi}H^3\sim \mathcal{O}(10^{-26})$ and $\ddot{\xi}H^2\sim
\mathcal{O}(10^{-28})$. In Table \ref{table1} we collect the
characteristic features of this model for convenience.
\begin{table}[h!]
  \begin{center}
    \caption{\emph{\textbf{Several Viable Einstein-Gauss-Bonnet Models with $\xi'(\phi)=\frac{\lambda  V'(\phi )}{V(\phi )^2}$.}}}
    \label{table1}
    \begin{tabular}{|r|r|r|r|}
     \hline
      \textbf{Model}   & \textbf{Planck Constraints} & \textbf{Values of Scalar Field} &
      \textbf{Gravitational Wave Speed}
      \\  \hline
      $V(\phi)=\frac{M (\kappa  \phi )^2}{d+(\kappa  \phi )^2}$ & $(d,\lambda,M)=(3,10^{-13},2.583\times
10^{-10})$ & $(\phi_i,\phi_f)=(5.64629,1.03964)$ &
$c_T^2-1=-1.19917\times 10^{-16}$
\\  \hline
 $V(\phi)=M \left(1-\frac{\delta }{\kappa  \phi ^2}\right)$ & $(\delta,\lambda,M)=(10^{-3},10^{-13},4.2706\times 10^{-12})$ & $(\phi_i,\phi_f)=(0.787,0.115)$ &
$c_T^2-1=-2.16654\times 10^{-17}$  \\  \hline
 $V(\phi)=M \left(1-\frac{\delta }{\kappa  \phi }\right)^2$ & $(\delta,\lambda,M)=(3,10^{-13},1.8597\times 10^{-9})$ & $(\phi_i,\phi_f)=(11.385,4.048)$ &
$c_T^2-1=-5.39095\times 10^{-16}$  \\  \hline
    \end{tabular}
  \end{center}
\end{table}
Let us now consider the potential,
\begin{equation}\label{potential1a}
V(\phi)=M \left(1-\frac{\delta }{\kappa  \phi ^2}\right)\, ,
\end{equation}
which is known in the single scalar field theory literature as
brane inflation potential (BI), developed in Refs.
\cite{Lorenz:2007ze,Jones:2002cv,Alexander:2001ks,Burgess:2001fx,Pogosian:2003mz,Brandenberger:2007ca,Ma:2008rf}.
In this case, the derivative of the Gauss-Bonnet scalar coupling
function in natural units has the form,
\begin{equation}\label{xiphi1}
\xi'(\phi)=\frac{2 \delta  \lambda  \phi }{M \left(\delta -\phi
^2\right)^2}\, ,
\end{equation}
and the first slow-roll index becomes,
\begin{equation}\label{slowrollindexena1}
\epsilon_1=\frac{2 \delta ^2 (4 \lambda +3)^2}{9 \left(\phi
^3-\delta  \phi \right)^2}\, ,
\end{equation}
and therefore one may again solve the equation
$|\epsilon_1(\phi_f)|=1$ and obtain  the value $\phi_f$, which we
omit for brevity. In the case at hand, the integral involved in
the computation of the $e$-foldings number becomes,
\begin{equation}\label{integraln11}
N=-\frac{\frac{3 \delta  \phi ^2}{2}-\frac{3 \phi ^4}{4}}{8 \delta
\lambda +6 \delta }\Big{|}_{\phi_f}^{\phi_i}\, ,
\end{equation}
and accordingly, the value of the scalar field $\phi_i$ can be
obtained. In this case too, a large range of the values of the
free parameters can guarantee a viable inflationary phenomenology,
for example for $N=48$ and for
$(\delta,\lambda,M)=(10^{-3},10^{-13},4.2706\times 10^{-12})$, we
get, $n_{\mathcal{S}}=0.9687$, $r= 0.00026$ and
$n_{\mathcal{T}}=-0.000016$, therefore the model is viable
regarding the observational indices of inflation and well
compatible with the Planck data. This can also be seen in Fig.
\ref{plot2} where we plot the current model's predictions versus
the Planck 2018 likelihood curves, for $N=[46,52]$.
Regarding the amplitude of the scalar perturbations in this case,
for the choice $(\delta,\lambda,M)=(10^{-3},10^{-13},4.2706\times
10^{-12})$ we get exactly $\mathcal{P}_{\zeta}(k_*)=2.196\times
10^{-9}$, and as expected, the amplitude of the scalar
perturbations crucially depends on the value of $M$ in this case
too. For $(\delta,\lambda,M)=(10^{-3},10^{-13},4.2706\times
10^{-12})$ in this case, the scalar field values at the beginning
and the end of inflation are sub-Planckian, specifically, for
$(\delta,\lambda,M)=(10^{-3},10^{-13},4.2706\times 10^{-12})$ we
get, $(\phi_i,\phi_f)=(0.787,0.115)$. Regarding the gravitational
wave speed, in this case we have $c_T^2-1=-2.16654\times
10^{-17}$, so in this case too, the model is consistent with the
constraint (\ref{GWp9}). Regarding the constraints
(\ref{additionalconstraints}), these are well satisfied for this
model too, for example for
$(\delta,\lambda,M)=(10^{-3},10^{-13},4.2706\times 10^{-12})$ we
have $\dot{\phi}^2\sim\mathcal{O}(10^{-17})$, $\dot{\xi}H^3\sim
\mathcal{O}(10^{-29})$ and $\ddot{\xi}H^2\sim
\mathcal{O}(10^{-30})$. In Table \ref{table1} we collected all the
characteristic features for this model too. Let us now consider
another viable potential,
\begin{equation}\label{potential2}
V(\phi)=M \left(1-\frac{\delta }{\kappa  \phi }\right)^2\, ,
\end{equation}
which is not classified into known single scalar field potentials,
but provides an interesting viable phenomenology, and we call it
inverse power law (IPL). In this case, the derivative of the
Gauss-Bonnet scalar coupling function in natural units has the
form,
\begin{equation}\label{xiphi2}
\xi'(\phi)=-\frac{2 \delta  \lambda  \phi }{M (\delta -\phi )^3}\,
,
\end{equation}
and the first slow-roll index becomes in this case,
\begin{equation}\label{slowrollindexena2}
\epsilon_1=\frac{2 \delta ^2 (4 \lambda +3)^2}{9 \phi ^2 (\delta
-\phi )^2}\, ,
\end{equation}
and again one may solve the equation $|\epsilon_1(\phi_f)|=1$ and
obtain the value $\phi_f$, which we again omit for brevity. In
this case, the integral involved in the computation of the
$e$-foldings number becomes,
\begin{equation}\label{integraln12}
N=-\frac{3 \left(\frac{\delta  \phi ^2}{2}-\frac{\phi
^3}{3}\right)}{2 \delta  (4 \lambda
+3)}\Big{|}_{\phi_f}^{\phi_i}\, ,
\end{equation}
and accordingly, the value of the scalar field $\phi_i$ can be
obtained. In this scenario too, a large range of the values of the
free parameters guarantees a viable inflationary phenomenology,
for example for $N=48$ and if we choose
$(\delta,\lambda,M)=(3,10^{-13},1.8597\times 10^{-9})$, we get,
$n_{\mathcal{S}}=0.9686$, $r= 0.066$ and
$n_{\mathcal{T}}=-0.0041$, therefore the model is marginally
viable regarding the observational indices of inflation and well
compatible with the Planck data. This can also be seen in this
case in Fig. \ref{plot3} where we plot the current model's
predictions versus the Planck 2018 likelihood curves, for
$N=[46,52]$.
Now, regarding the amplitude of the scalar perturbations in this
case, for the choice $(\delta,\lambda,M)=(3,10^{-13},1.8597\times
10^{-9})$ we get exactly $\mathcal{P}_{\zeta}(k_*)=2.196\times
10^{-9}$. For $(\delta,\lambda,M)=(3,10^{-13},1.8597\times
10^{-9})$ in this case, the scalar field values at the beginning
and the end of inflation are super-Planckian, specifically, for
$(\delta,\lambda,M)=(3,10^{-13},1.8597\times 10^{-9})$ we get,
$(\phi_i,\phi_f)=(11.38,4.04)$, and this is a model dependent
feature, in other models we found, the scalar field values are
sub-Planckian. In all the cases we found, it is mentionable that
the scalar field values decrease as time evolves, which is an
anticipated feature, which we shall use in the next section.
Regarding the gravitational wave speed, in this case we have
$c_T^2-1=-5.3909\times 10^{-16}$, so in this case too, the model
is consistent with the constraint (\ref{GWp9}). Regarding the
constraints (\ref{additionalconstraints}), these are well
satisfied for this model too, for example if we choose
$(\delta,\lambda,M)=(3,10^{-13},1.8597\times 10^{-9})$ we get
$\dot{\phi}^2\sim\mathcal{O}(10^{-12})$, $\dot{\xi}H^3\sim
\mathcal{O}(10^{-27})$ and $\ddot{\xi}H^2\sim
\mathcal{O}(10^{-35})$. In Table \ref{table1} we collected all the
characteristic features for this model too.

In conclusion, we presented a variety of models which provide a
viable inflationary era, in addition to the compatibility with all
the constraints coming from GW170817 and all the additional
required constraints. Now in the next section we shall consider an
important issue having to do with the functional form of the
Gauss-Bonnet coupling function during the reheating era. In this
section we chose arbitrarily the scalar Gauss-Bonnet coupling,
however this is not the case in the reheating era, in which case
the functional form of the scalar Gauss-Bonnet coupling is related
to the form of the given potential. An outcome of this section
that will be used in the next section, is the well anticipated
fact that the scalar field values decrease as the cosmic time
evolves.

\section{Behavior of the Einstein-Gauss-Bonnet Gravity During the Reheating Era}

\begin{figure}
\centering
\includegraphics[width=20pc]{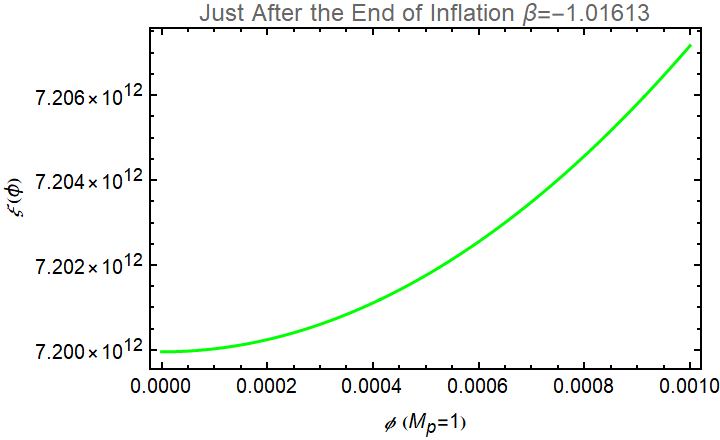}
\includegraphics[width=20pc]{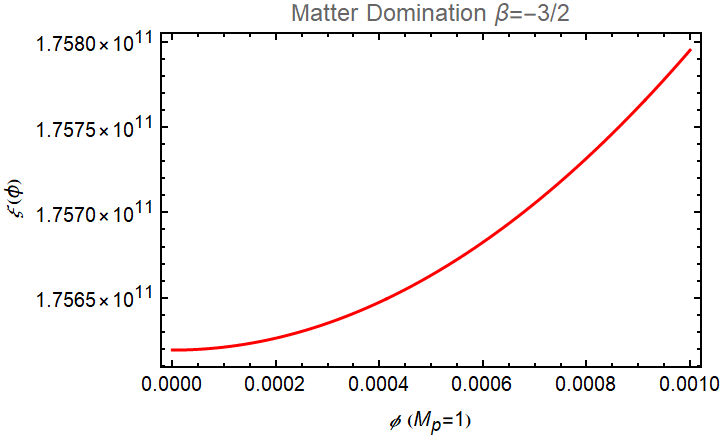}
\includegraphics[width=20pc]{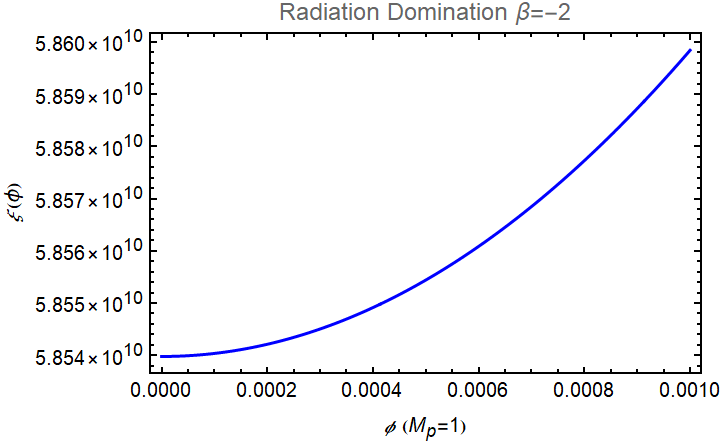}
\caption{The scalar Gauss-Bonnet coupling function $\xi(\phi)$
during the reheating era as a function of the scalar field in
Planck units, for the scalar potential (\ref{potential1a}). We
used three distinct values for $\beta$, namely $\beta=-1.01613$
which corresponds to the green curve (nearly the end of inflation
era value of the total EoS $w_{eff}=-1/3$), $\beta=-3/2$ (matter
domination era and red curve) and $\beta=-2$ (radiation domination
era and blue curve).}\label{plot4}
\end{figure}

\begin{figure}
\centering
\includegraphics[width=20pc]{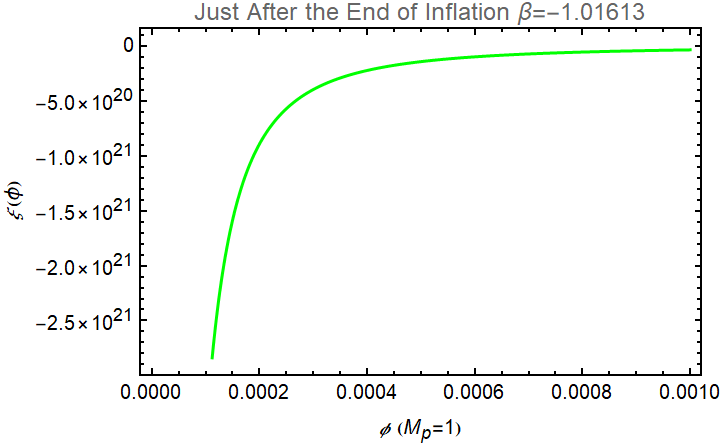}
\includegraphics[width=20pc]{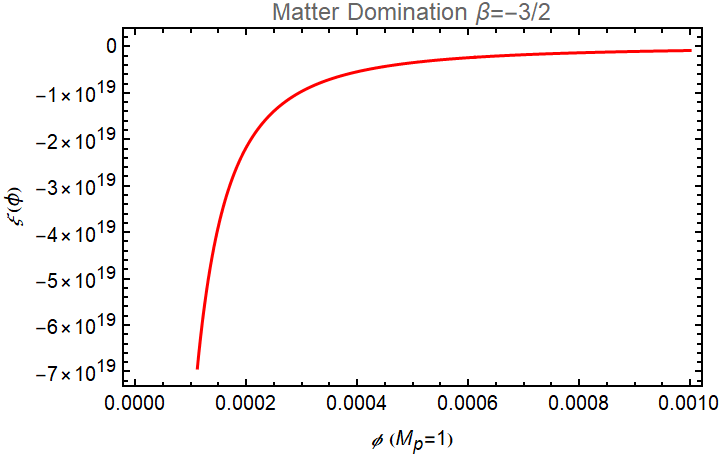}
\includegraphics[width=20pc]{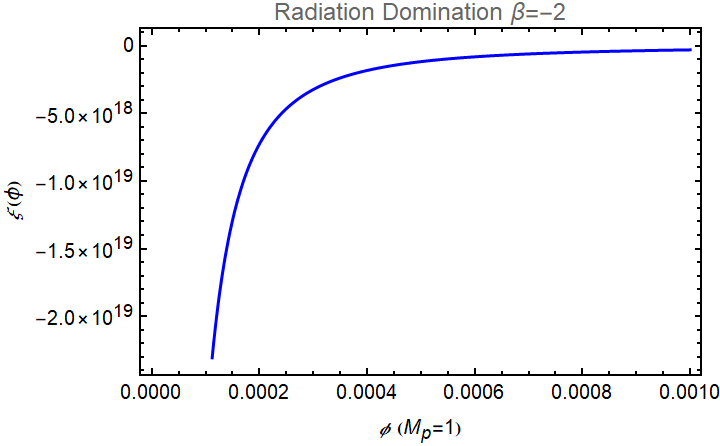}
\caption{The scalar Gauss-Bonnet coupling function $\xi(\phi)$
during the reheating era as a function of the scalar field in
Planck units, for the scalar potential (\ref{potential}). We used
three distinct values for $\beta$, namely $\beta=-1.01613$ which
corresponds to the green curve (nearly the end of inflation era
value of the total EoS $w_{eff}=-1/3$), $\beta=-3/2$ (matter
domination era and red curve) and $\beta=-2$ (radiation domination
era and blue curve). }\label{plot5}
\end{figure}

In this section we shall investigate the reheating era for the
Einstein-Gauss-Bonnet gravity framework we developed in the
previous sections, and we shall demonstrate that the scalar
Gauss-Bonnet coupling is constrained to be related to the scalar
potential. The scale factor of the Universe during the reheating
era is assumed to have the general form  $a(t)\sim t^{-1/\beta}$
(assuming that $\beta<0$) which results to the relation between
the Hubble rate and its time derivative $\dot{H}=\beta H^2$. Now
the reason for this is simple, the total equation of state (EoS)
parameter $w_{eff}$, defined as,
\begin{equation}\label{weff}
w_{eff}=-1-\frac{2\dot{H}}{3H^2}\, ,
\end{equation}
during the reheating era takes values in the range
$-\frac{1}{3}\leq w_{eff}\leq \frac{1}{3}$, which is explained
simply because the value $w_{eff}=-\frac{1}{3}$ corresponds to the
non-acceleration value exactly at the end of the inflationary era,
while $w_{eff}=\frac{1}{3}$ corresponds to the value of a
radiation domination era. The reheating era though is a vague era
for which little are known thus it is far from certain what was
the total EoS parameter during that era. We will thus assume that
during the reheating era, the total EoS parameter takes various
characteristic values, varying from the value nearly at the end of
the inflationary era, $w_{eff}=-1/3.1$ to the value during the
radiation domination era $w_{eff}=1/3$. Specifically, we assume
that $w_{eff}=-1/3.1$ which corresponds to $\beta=-1.01613$ (which
is slightly smaller than the end of inflation era value of the EoS
$w_{eff}=-1/3$), $w_{eff}=0$ which corresponds to $\beta=-3/2$ and
describes a matter domination era, and finally $w_{eff}=1/3$ which
corresponds to $\beta=-2$ and describes a radiation domination
era. One can easily work out this scenario to find the behavior of
the Gauss-Bonnet scalar coupling function $\xi(\phi)$, assuming
again that the inequalities of Eq.
(\ref{actualgw170817constraints}) hold true, so after some algebra
one gets,
\begin{equation}\label{r1}
H^2\simeq \frac{\kappa^2 V}{3+\beta}\, ,
\end{equation}
\begin{equation}\label{r2}
\ddot{\phi}\simeq -\frac{\beta}{3+\beta}V'\, ,
\end{equation}
\begin{equation}\label{r3}
\dot{\phi}^2\simeq -\frac{2\beta}{3+\beta}V\, ,
\end{equation}
and finally,
\begin{equation}\label{r4}
\xi'(\phi)\simeq \frac{-\frac{3}{3+\beta}V'+3\gamma \delta
V}{\mathcal{A}V^2}\, ,
\end{equation}
with,
\begin{equation}\label{r5}
\gamma=\sqrt{-\frac{2\beta}{3+\beta}},\,\,\,\delta=\sqrt{\frac{\kappa^2}{3+\beta}},\,\,\,\mathcal{A}=\frac{12(\beta+1)\kappa^4}{(3+\beta)^2}\,
.
\end{equation}
One can solve the differential equation (\ref{r4}), given the
scalar field potential, and find the function $\xi(\phi)$ or can
have directly $\xi'(\phi)$ and investigate the behavior of it as a
function of the scalar field. We shall reveal the behavior of
$\xi(\phi)$ for the three values of $\beta$ we mentioned earlier
and for some characteristic viable potentials we presented in the
previous section, using the corresponding values of the free
parameters in order to see the behavior of $\xi (\phi)$ during the
reheating era. Consider first the BI potential (\ref{potential1a})
so by substituting the potential in the differential equation
(\ref{r4}) we obtain the following solution for the scalar
Gauss-Bonnet coupling function,
\begin{equation}\label{finalexpsolution1}
\xi(\phi)=\frac{\beta  \left(-\frac{\sqrt{-\frac{\beta }{\beta
+3}} \delta }{\phi ^2-\delta }-\sqrt{2} \beta  \sqrt{\frac{\delta
}{\beta +3}} \tanh ^{-1}\left(\frac{\phi }{\sqrt{\delta
}}\right)+\sqrt{2} \beta  \sqrt{\frac{1}{\beta +3}} \phi
\right)}{4 (\beta +1) \left(-\frac{\beta }{\beta +3}\right)^{3/2}
M}\, .
\end{equation}
Using $(\delta,\lambda,M)=(10^{-3},10^{-13},4.2706\times
10^{-12})$ and working again in Planck units, we can see the
behavior of the Gauss-Bonnet coupling function
(\ref{finalexpsolution1}) in Fig. \ref{plot4}, for
$\beta=-1.01613$ which corresponds to the green curve (nearly the
end of inflation era value of the total EoS $w_{eff}=-1/3$),
$\beta=-3/2$ (matter domination era and red curve) and $\beta=-2$
(radiation domination era and blue curve). As it can be seen, as
the scalar field values increase, the scalar Gauss-Bonnet scalar
coupling decreases, for all the aforementioned cases, the
difference is the magnitude of the function $\xi (\phi)$ for the
three distinct scenarios.
Now let us consider the RDI potential (\ref{potential}) so by
substituting the potential in the differential equation (\ref{r4})
in this case we obtain the following solution,
\begin{equation}\label{finalexpsolution12}
\xi(\phi)=\frac{(\beta +3) \left(\frac{\sqrt{2} (\beta +3)
\sqrt{-\beta } \phi ^3}{\beta +3}-\frac{\sqrt{2} \sqrt{-\beta }
(\beta +3) d \phi }{\beta +3}+d\right)}{4 (\beta +1) M \phi ^2}\,
.
\end{equation}
Using $(d,\lambda,M)=(3,10^{-13},2.583\times 10^{-10})$ in Planck
units, we plot the Gauss-Bonnet coupling function
(\ref{finalexpsolution12}) as a function of the scalar field in
Fig. \ref{plot5}. Contrary to the previous model, in this case the
scalar Gauss-Bonnet coupling function increases as the scalar
field values increase. The behavior for the three distinct values
of $\beta$ we used is qualitatively similar and the only
difference among the three cases for the potential under study is,
as in the previous case, the magnitude of the Gauss-Bonnet
coupling function.
In conclusion, the behavior of the scalar Gauss-Bonnet coupling
during the reheating era for the Einstein-Gauss-Bonnet models we
considered in this paper strongly depends on the scalar potential.
Thus it cannot be predicted how $\xi (\phi)$ will behave and its
behavior is more or less model dependent. Now a crucial comment
would be, whether the constrained form of the scalar Gauss-Bonnet
coupling function obtained during the reheating, can be the same
during the inflationary era. This is hard to decide because this
constraint was obtained for a constant-roll-like evolution of the
form $\dot{H}=\beta H^2$, and this would imply that the
inflationary era would be a power-law evolution. The whole
formalism should change though and we did not address this issue
in this paper, since it would be out of context. We aim to address
this research topic in a future work though.

\section*{Concluding Remarks}

In this work we developed the inflationary framework of
Einstein-Gauss-Bonnet theories that produce a propagating speed of
tensor perturbations that respect the constraint imposed by the
GW170817 event, namely $\left| c_T^2 - 1 \right| < 6 \times
10^{-15}$ in natural units. In general, in these theories the
scalar Gauss-Bonnet coupling function $\xi (\phi)$ can be chosen
freely and there is no imposed direct relation between the
$\xi(\phi)$ and the scalar potential. We developed the
inflationary formalism assuming only a slow-roll era for the
scalar field and also imposing the GW170817 constraints on the
propagating speed of tensor perturbations which constrain the
scalar Gauss-Bonnet coupling function $\xi(\phi)$. We applied the
formalism using an interesting class of models and for several
potentials that produce a viable inflationary era compatible with
the Planck constraints and also respect the GW170817 constraints.
Also we considered the reheating era in the context of
Einstein-Gauss-Bonnet gravitational theories, in which case the
Hubble rate obeys a constant-roll-like evolution relation
$\dot{H}=\delta H^2$. Assuming that the GW170817 constraints also
hold true, it proves that the scalar Gauss-Bonnet coupling
function $\xi(\phi)$ and the scalar potential obey a differential
equation. Therefore, given the potential, one finds the scalar
Gauss-Bonnet coupling function, which is different from the one
during the inflationary era. We considered two examples and we
examined the behavior of the scalar Gauss-Bonnet coupling function
during the reheating era. An issue which we did not address is to
examine the inflationary phenomenology of Einstein-Gauss-Bonnet
theories for which the Hubble rate obeys a constant-roll-like
evolution during the inflationary era. This would produce a
power-law inflationary era, thus the formalism and field equations
of the corresponding Einstein-Gauss-Bonnet theory would be
different compared to the one we developed here. There is strong
motivation for such constant-roll inflationary conditions, which
were firstly introduced in \cite{Martin:2012pe}, for example
primordial black hole formation and induced gravitational waves
\cite{Motohashi:2019rhu,Tomberg:2023kli}. We aim to address this
constant-roll inflationary era research task in a future work.

\section*{Acknowledgments}

This research has been is funded by the Committee of Science of
the Ministry of Education and Science of the Republic of
Kazakhstan (Grant No. AP14869238).

\end{document}